\documentclass[12pt]{article}
\usepackage{graphicx}

\usepackage{makeidx}

\usepackage{amssymb}

 \newcommand{\kr}{k_{\rho}}

 \newcommand{\mrm}{{\rm m}}

 \newcommand{\minrm}{{\rm min}}

 \newcommand{\maxrm}{{\rm max}}

 \newcommand{\text}{\rm}

 \newcommand{\ug}{ \; = \; }

 \newcommand{\de}{\delta}
 
 \newcommand{\bb}{\begin{equation}}
 \newcommand{\ee}{\end{equation}}
 \newcommand{\bc}{\begin{center}}
 \newcommand{\ec}{\end{center}}
 \newcommand{\bega}{\begin{eqnarray}}
 \newcommand{\ega}{\end{eqnarray}}
 \newcommand{\begae}{\begin{eqnarray*}}
 \newcommand{\egae}{\end{eqnarray*}}

 \newcommand{\h}{\hspace*{4ex}}

 \newcommand{\om}{\omega}

 \newcommand{\cent}{\centerline}
 \newcommand{\vs}{\vspace*}

\begin{document}
\baselineskip 0.5cm

\begin{center}

{\large {\bf Parabolic antennas, and circular slit arrays, for the generation of
Non-Diffracting Beams of Microwaves$^{\: (\dag)}$ }} \footnotetext{ $^{\: (\dag)}$
E-mail addresses: \ recami@mi.infn.it ; \ mzamboni@decom.fee.unicamp.br }

\end{center}

\vs{5mm}

\cent{ Michel Zamboni-Rached$^{\; 1,2}$ \ and \ Erasmo Recami$^{\; 2,3,4}$ }

\vs{0.2 cm}

\centerline{$^{1}$ {\em Photonics Group, Electrical \& Computer Engineering, University of Toronto, CA}}
\centerline{$^{2}$ {\em DECOM, FEEC, Universidade Estadual de Campinas
(UNICAMP), Campinas, SP, Brazil}}
\cent{$^{3 \;}$ {\em Facolt\`a di Ingegneria, Universit\`a statale di
Bergamo, Bergamo, Italy.}}
\cent{$^{4 \;}$ {\em INFN---Sezione di Milano, Milan, Italy.}}

\vs{0.5 cm}


\printindex

{\bf Abstract  \ --} \  We propose in detail Antennas for generating Non-Diffracting Beams of Microwaves,
for instance with frequencies of the order of 10 GHz, obtaining fair results even when having recourse
to realistic apertures endowed with reasonable diameters.  Our first proposal refers mainly to sets of
suitable annular slits, having in mind various possible applications, including remote sensing. \ Our
second proposal ---which constitutes one of the main aims of this paper--- refers to the alternative,
rather simple, use of a Parabolic Reflector, illuminated by a spherical wave source located on the paraboloid axis
but slightly displaced with respect to the Focus of the Paraboloid. Such a parabolic reflector yields
{\em extended focus} (non-diffracting) beams. [OCIS codes: 999.9999; 070.7545; 050.1120; 280.0280; 050.1755; 
070.0070; 200.0200.  Keywords: Non-Diffracting Waves; Microwaves; Remote sensing; Annular Arrays; Bessel beams; 
Extended focus; Reflecting paraboloids; Parabolic reflectors; Parabolic antennas].

\

\section{Introduction}

Since several years it has been discovered that also linear equations like the ordinary wave equations (scalar, vectorial, spinorial...) admit of “soliton-like” solutions,  known as Localized Waves (LW) \cite{Livro1} but that more properly ought to be called Non-diffracting Waves (NDW) \cite{Livro2}. Actually, they possess peculiar properties, as the one of resisting diffraction over long field-depths, and of self-reconstructing themselves after obstacles with size of the order of the antenna's (and not of their wavelength's).

\h They are more suited than the Gaussian waves to represent even elementary particles; and indeed “localized solutions” exist not only to the K-G or Dirac equations, but also ---mutatis mutandis--- to the Schroedinger equation (and even to the Einstein equations of GR).

\h Theory, and applications, of the Non-diffracting Waves have been developed and realized in Acoustics, and even more in Electromagnetism and Optics, during the last two decades, and in particular in recent years.  For instance, our first book (J.Wiley, 2008) on NDWs collected the contributions from 10 research groups [its Contents can be downloaded from the site www.unibg.it/recami], while, to our second book (J.Wiley, 2014) on NDWs, 20 research groups contributed. \ Indeed, Non-Diffracting Waves are continuously having, and promising, more and more applications. A further interesting fact, a priori, is that their
peak-velocities can run from zero to infinity, as it was theoretically derived, in terms of exact analytic solutions, for instance from Maxwell equations only, confirmed by numerical evaluations, and experimentally verified (in the case of optics and microwaves, interesting experimental papers started to appear in the {\em PRL} of 1997, while
in Acoustics they had started to appear in 1992).  But, as we were saying, the NDWs are important for their properties, independently of their peak velocity!  \ Nevertheless, many people have studied (theoretically, mathematically, and experimentally) the particular NDWs known as ``superluminal" X-shaped pulses, it being easier their theoretical and experimental construction.  But we later succeeded in studying also the ``more orthodox" subluminal solutions, even if expressed in terms of finite-limit integrals, more difficult to be analytically evaluated; and we have in particular discovered how to describe NDWs at rest: i.e., with a static envelope.  Such ``Frozen Waves" (FW) are expected to have even more incredible applications\cite{Patent}:  e.g., in medicine (for tumor cell destruction, without affecting the front and rear, or surrounding, tissues); or for new types of optical (or acoustic) tweezers; for particle guiding, etc.  The FWs have been experimentally produced in Optics in recent times, and in Acoustics (by simulated experiments, this time) even more recently.

\h To be a little more specific, the NDWs have become a hot topic nowadays in a variety of fields. \ In particular, their use, replacing laser beams for achieving multiple traps, has found many potential
applications in medicine and biomedicine (see, for instance, Refs.\cite{Arlt,Herman,Garces,ref4,ref5}). Even though their simple, multi-ringed structures are not always suitable enough [e.g., they aren't fit for an effective three-dimensional trap when single
beam setups are employed], nevertheless, with today techniques for their generation and real-time control,
non-diffracting beams have become ---better then focused Gaussian beams or others--- indispensable
``laser-type" beams for biological studies by means of optical tweezing and micromanipulation techniques. And elsewhere, in
fact, the theoretical aspects of an application, for example, in biomedical optics have been presented: Namely,
that of Optical Tweezers construction, and of Micro-manipulations, by NDWs in connection with the Generalized Lorenz-Mie
Theory\cite{IntroductionII}.

\h In this paper we propose in detail Antennas for generating Non-Diffracting Beams of Microwaves,
for instance with frequencies of the order of 10 GHz, obtaining fair results even when having recourse
to realistic apertures possessing reasonable diameters.  Our first proposal refers mainly to sets of
suitable annular slits, having in mind various possible applications, including remote sensing. \ Our
second proposal ---which constitutes one of the main aims of this paper--- refers to the rather simple,
alternative use of a Parabolic Reflector, illuminated by a spherical wave source located on the paraboloid axis
but slightly displaced with respect to the Focus of the Paraboloid. Such a parabolic reflector yields
(non-diffracting) beams with an {\em extended focus}. \ The present paper reports about work performed by us
especially in 2011, and 2012.


\subsection{Another brief preamble}

Before going on, let us recall ---for the readers not well acquainted with the topic--- that even finite-energy NDWs
(Bessel beams, e.g.), obtained for example by suitable truncations, still keep their extraordinary properties all along their
depth of field, {\em much} longer that the one possessed by diffracting waves (like the gaussian ones). \ Incidentally, the
field-depth is as long as the ``extended-focus" of the NDW.

\begin{figure}[!h]
\begin{center}
 \scalebox{2.5}{\includegraphics{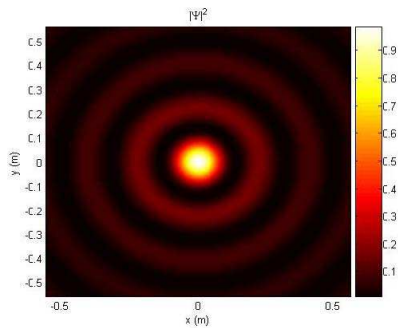}}
\end{center}
\caption{(Color online) Consider a Bessel beam with initial spot size $r_0 = 9$ cm, frequency $f$ of 15 GHz, and generated by a finite antenna with radius $R = 0.56$ m.  Its axicon angle\cite{Livro1} is of 0.055 rad, while its field depth will is $Z =10.4$ m (that is, the beam maintains its resolution ---spot size and intensity--- all along its extended focus). This figure shows the form of its field intensity in {\em any} transverse plane in the range from $z=0$ to $z=10.4$ m.}
\label{FiguraUno}
\end{figure}

\h Consider for instance a frequency $f$ of 15 GHz, and a finite antenna with radius $R = 0.56$ m; and
suppose we wish a Bessel beam with initial spot size $r_0 = 9$ cm. We have then to use an axicon angle\cite{Livro1}
of 0.055 rad. The field depth, $Z$, of the Bessel beam will be 10.4 m: and the beam will maintain its resolution (spot size and
intensity) along all its extended focus (from 0 to 10.4 m). The form of the field intensity in any transverse plane in this range, that is from $z=0$ to about $z=10$ m, is shown in Figure \ref{FiguraUno}.

Instead, a gaussian beam with the same initial spot-size, will get the larger spot size of $0.19$ m after 5 m of propagation; and
$0.34$ m, even larger, after 10 m of propagation;  besides a substantial reduction of the spot intensity.

\

\begin{figure}[!h]
\begin{center}
 \scalebox{2.5}{\includegraphics{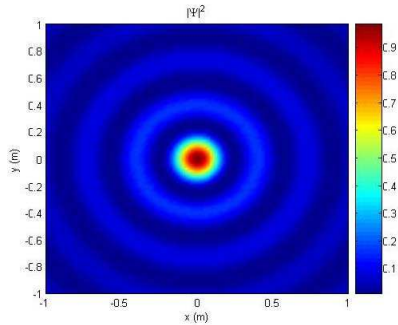}}
\end{center}
\caption{(Color online) As a second case, consider a Bessel beam with initial spot size $r_0 = 17.7$ cm, frequency $f$ of 15 GHz, and generated by a finite antenna with the larger radius $R = 1$ m.  Its axicon angle\cite{Livro1} is of 0.031 rad, while its field depth will arrive at $Z =10.4$ m
(that is, the beam maintains its resolution ---spot size and intensity--- all along its extended focus). This figure shows the form of its field intensity in {\em any} transverse plane in the range from $z=0$ to $z=33$ m.}
\label{FiguraDue}
\end{figure}

\h As a second case, consider a larger aperture radius, $R = 1$ m: a Bessel beam with the initial spot-size $r_0 = 17.7$ cm (which correspond to an axicon angle of 0.031 rad) will possess a field depth $Z = 33$ m, maintaining its spot size and intensity till the distance of 33 m. The form of the field intensity in any transverse plane in this range, from $z=0$ to $z=33$ m, is shown in Figure \ref{FiguraDue}.

\

Instead, after a distance of 33 m, a gaussian beam with the same initial spot-size, will get a quite larger
spot-size of $0.62$ m, besides a substantial reduction of its spot intensity.

\section{Antennas Generating Non-diffracting Beams\\
 (for remote sensing purposes, {\em et alia})} 

In our Ref.\cite{MRB}, we already considered the production of truncated pulses, having in mind ---among their applications---
also remote sensing.

\h Still for remote sensing {\em et alia}, let us here consider further possible antennas, suitable for the generation of
non-diffracting {\em beams} of microwaves; and choose e.g. the frequency of 15 GHz (and about 1 m for the aperture diameter). \ Such an antenna can simply be an array
of annular slits\cite{MRBunpub1,Antenna3Archives,mrb2,PatentII}. \ As we were saying, we shall propose later on the alternative use of a parabolic reflector, illuminated by a
spherical wave source located on the paraboloid axis but slightly displaced w.r.t. the reflector
focus\cite{MRBunpub2,MRunpub}.

\h The said frequency corresponds to a $\lambda$ of 2 cm, and
aperture radii {\em much} larger than $\lambda$ would be needed
for creating highly efficient non-diffracting beams, with spots
of the order of $\lambda$ and with a quite large field-depth. \

\h However, we are going to make a more efficient choice from the realistic point of view.

\h Let us base ourselves on the scalar approximation; and recall that a Bessel beam (Bb)
with axial symmetry can be written

\bb \psi(\rho,z,t) = J_0(\kr\rho)\exp{i(k_z z - \om t)} \; ,
\label{bb}\ee

which refers to an ideal beam endowed with an infinite depth of field,
that is, with an invariable transverse structure and with a spot given
by $\Delta\rho = 2.4/\kr$ at any positions of its. We know that such a Bb
would be associated with an infinite power flux through a transverse
surface as the $z=0$ one. One needs truncating it by a finite aperture
with a radius $R >> \Delta\rho$, and it gets the finite field-depth
$Z = R/\tan(\theta)$, where the Bb axicon angle $\theta$ depends on the
longitudinal and transverse wavenumbers through the relations
$k_z = \om/c\,\cos(\theta)$ \ and \ $\kr = \om/c\,\sin(\theta)$. \
Simple geometric optics reasonings tell us that, in the region $0<z<Z$ and $0<\rho<(Z-z)\tan(\theta)$,
the truncated Bb can still be described by the ideal solution (\ref{bb}). \ However,
when the aperture radius does not obey the relation $R>>\Delta\rho$, one normally has
to resort to lengthy numerical simulations, based on the diffraction integrals, for
obtaining the field emanated by the finite antenna. And this is just our case.

\begin{figure}[!h]
\begin{center}
 \scalebox{2.5}{\includegraphics{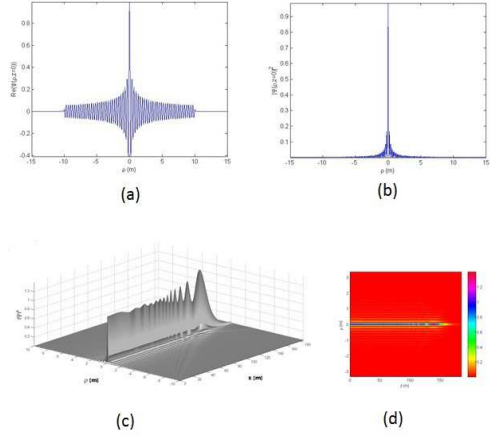}}
\end{center}
\caption{(Color online) Figures referring to a Bb with axicon angle $\theta = 0.062$ rad, and frequency 15 GHz
(therefore with spot $\Delta\rho=12$ cm), truncated by a finite circular aperture having the {\em large} radius
$R=10$ m. \ One expects the emanated field to be approximately given by Eq.(\ref{bb}) in the region
$0<z<Z$ and $0<\rho<(Z-z)\tan(\theta)$, with $Z = 161.1$ m. \ For comparison, see, by contrast, the case in the next Figure.}
\label{figure1}
\end{figure}

\h {\em However,} we have at our disposal the method in Ref.\cite{MRB} which yields analytic expressions for truncated
fields, allowing us to get our results in a few seconds. We already applied it even to antennas composed of
annular slits. \ Before going on, let us recall also some characteristics of a truncated Bessel beam. \ Consider
a Bb with axicon angle $\theta = 0.062$ rad, and frequency 15 GHz
(therefore with spot $\Delta\rho=12$ cm), truncated by a finite circular aperture having radius
$R=10$ m. \ One expects the emanated field to be approximately given by Eq.(\ref{bb}) in the region
$0<z<Z$ and $0<\rho<(Z-z)\tan(\theta)$, with $Z = 161.1$ m. \ [For the Bb, we are calling spot-radius the distance
(in the transverse direction, starting from $\rho=0$) at which one meets the first zero of the field intensity].
The Figures \ref{figure1} show:  the field at the aperture (figure (a)) and its intensity (figure (b)), as well as the 3D intensity of the emanated field (figure (c)) and its projection (figure (d)).

\begin{figure}[!h]
\begin{center}
 \scalebox{2.7}{\includegraphics{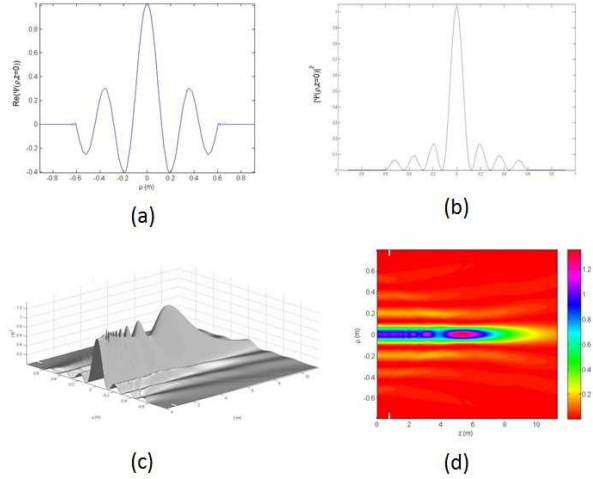}}
\end{center}
\caption{(Color online) Figures showing the behavior of a Bessel beam (Bb) truncated by a finite aperture too small for an efficient
non-diffracting beam (see the text): Indeed, only 3 lateral intensity rings survive the truncation.
} \label{figure2}
\end{figure}

\h If the Bb is truncated by a much smaller aperture (with radius $R=61$ cm) one cannot use any longer for the field-depth
the expression $R/\tan(\theta)$ which would yield $Z=9.8$ m. \ From the Figures \ref{figure2} one can see that, on the
contrary, the field starts suffering a strong decay at a lower distance, $z \simeq 6$ m; and that the Bb lateral intensity rings
(only 3 in this case) start to deteriorate even earlier: Since the intensity rings, too few, are unable to reconstruct the central
spot at the (large) distance $Z$.

\begin{figure}[!h]
\begin{center}
 \scalebox{2.22}{\includegraphics{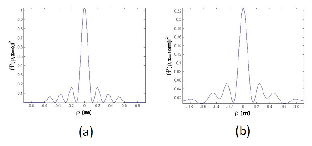}}
\end{center}
\caption{(Color online) Comparison of the transverse shapes of a Bb, truncated by an aperture $R=61$ cm:  \ (a) in the aperture plane, that is, $z=0$;  \ and  \ (b) after $10$ m of propagation, that is, at $z=10$ m.} \label{figure3}
\end{figure}

\h Nevertheless, even if the Bb starts decaying in this case at $z=6$ m, that is, well before $Z=R/\tan(\theta)=9.8$ m, its
spot width keeps its value for larger distances. \ Figures \ref{figure3} depict the Bb transverse intensity at $z=0$ and
after $10$ m of propagation: That is, at $z=10$ m.  One can see that the spot intensity decays at $1/4$ of its initial
value, but its radius changes only a little, from about $\Delta\rho(z=0)=12$ cm  \ to \ $\Delta\rho(z=10 \; \mrm)=15$ cm.

\begin{figure}[!h]
\begin{center}
 \scalebox{2.42}{\includegraphics{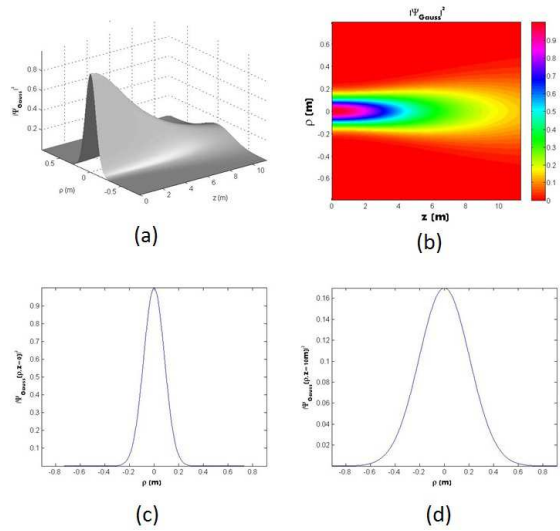}}
\end{center}
\caption{(Color online) Figure showing the evolution of a {\em gaussian} beam with the initial spot radius $\Delta\rho=12$ cm.}
\label{figure4}
\end{figure}

As we can expect, a {\em gaussian beam} with initial spot-radius $\Delta\rho(z=0)=12$ cm would double
such a width already at $3.9$ m, while at $z=10$ m its spot would have a central intensity almost 6 times smaller
than the initial one and, even worse, a tripled radius ($\Delta\rho(z=10 \; \mrm)=30$ cm).  \ This is shown by the
Figures \ref{figure4}.

All what precedes verifies that, even if the last {\em Bessel} beam, considered in Figures \ref{figure2},
is so severely truncated as to remain with only 3 of its lateral intensity rings, nevertheless it is still
able to keep its spot spatial shape (even if not its intensity) for distances much larger than for a gaussian
beam.

\

{\em Let us go on to our Prototypes} (cf. also our second Patent \cite{PatentII}).

\

\subsection{Antennas composed by annular slits}

Let us first consider\cite{MRBunpub1,Antenna3Archives,mrb2,PatentII} a circular
aperture surrounded by a set of concentric annular slits, aiming at producing by such an array (in
an approximate way) a Bb with axicon angle $\theta = 0.062$ rad, frequency $15$ GHz
(with a spot, therefore, $\Delta\rho=12$ cm), truncated by a circular aperture having $R=61$ cm.
A simple possibility is modeling the array of annular slits (plus the central aperture), and their
excitations, just taking in mind the shape itself of the desired (truncated) Bb: That is to say, we
can put the slits between the consecutive zeros of the Bessel function, and illuminate them by
uniform fields whose amplitude varies (passing from one slit to the other) according to the maximum
magnitude of the Bessel function in the corresponding intervals.

\begin{figure}[!h]
\begin{center}
 \scalebox{4.3}{\includegraphics{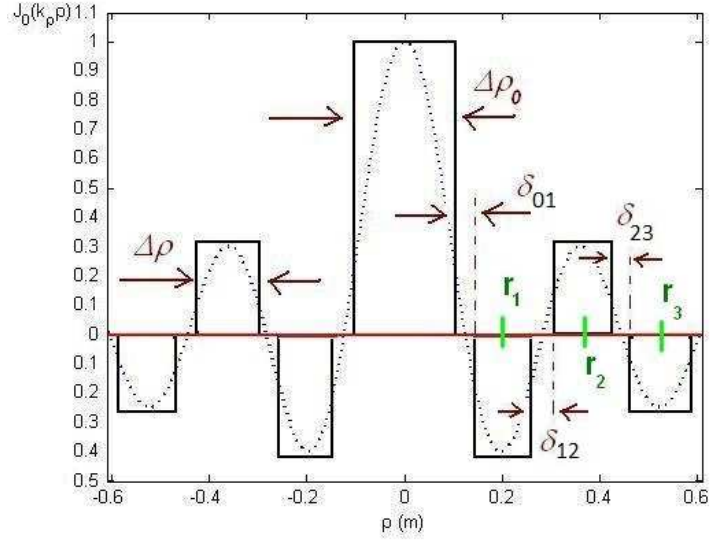}}
\end{center}
\caption{(Color online) This figure shows the spatial modeling of our first array of annular slits, and their excitations;
on the basis of the desired (truncated) Bb, at $z=0$. The dotted curves refer to the chosen Bessel
function, while the steps indicate location and widths of the slits, as well as the uniform field
amplitudes to be applied to each of them.}
\label{figure5}
\end{figure}

\h Figure \ref{figure5} is self-explicative.

It refers to our first Prototype. \ The values of the parameters are:

\

{\bf Prototype no. 1:} --- $r_1 = \pi(1+1/4)/\kr = 0.20$ m, \ $r_2 = \pi(2+1/4)/\kr =
0.36$ m, \ $r_3 = \pi(3+1/4)/\kr = 0.52$ m, \ $\Delta\rho_0
= 0.23$ m, \ $\Delta\rho_1=\Delta\rho_2=\Delta\rho_2=\Delta\rho= 0.13$ m, \
$\delta_{01}=0.021$ m, \ and \  $\delta_{12}=\delta_{23}=0.034$ m. \ \
The numerical values of the uniform field amplitudes within each slit are
given by the peak values assumed by the Bessel function therein.

\

Notice that the amplitude values pass from positive to negative values, and viceversa,
at each change of slit: This has to be strictly obeyed. \ The numerical values of the
field in the slits\footnote{Here $\Psi_n$ is the field numerical value in the $n$-th slit, where
$n=0$ refers to the central circular aperture (that can be called the slit number zero). Analogously,
$r_n$ is the radius of the $n$-th slit, while the radius of the central circle is called
$\Delta\rho_0$. One should not forget that \ $\kr = (\om/c) \sin (\theta) =  19.46 \ \mrm^{-1}$. }
are: \ $\Psi_0 = 1$ a.u., \ $\Psi_1 = J_0(\kr r_1)
= -0.4026$ a.u., \ $\Psi_2 = J_0(\kr r_2) = 0.3001$ a.u., \ $\Psi_3 = J_0(\kr r_3) =
-0.2497$ a.u.

\begin{figure}[!h]
\begin{center}
 \scalebox{2.1}{\includegraphics{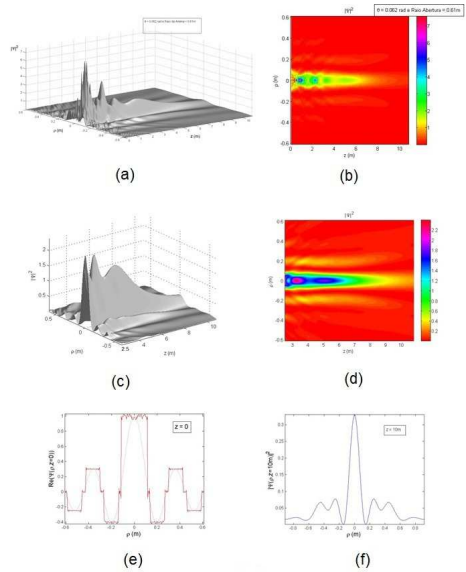}}
\end{center}
\caption{(Color online) Field emanated by Prototype 1. \ See the text.} \label{figure6}
\end{figure}

\h Figures \ref{figure6} show the field emanated by such an antenna. \ In figures (a) and (b)
the beam does not resemble a truncated Bb, because near the aperture the field is dominated by
some isolated intensity peaks (due to the slit edges themselves), which make the
field blurred. \ However, figures (c) and (d) present the field after $z=2.5$ m, and they do show similarity
with a truncated Bb. \ In figure (e), the red color (Color online) describes the field at
$z=0$; while the dotted line shows the Bessel function ``discretized" by the assumed field uniformity
inside each slit. Notice that this figure describes the real part of the
beam, with its positive or negative values of the amplitude: Such values are to be exactly
reproduced in the apparatus for the beam generation. \ At last, figure (f) shows the beam transverse intensity shape
after 10 m of propagation; that is, at $z=10$ m.  Notwithstanding the intensity drop (which decreases to 1/3 of the
one at $z=0$), the spot radius varies very little.

\

{\bf Prototype no. 2:} --- The spatial structure of the antenna is left unchanged, and one modifies
only the value of the uniform fields illuminating the slits. Simply, the uniform field at the circular
central aperture is not changed, while inside the slits ($n>1$) the value of each $\Psi_n$ is multiplied
by $\sqrt{n+1}$. \ Therefore:

$\Psi_0 = 1$, \ $\Psi_1 = \sqrt{2}\;J_0(\kr r_1)=-0.57$ a.u., \
$\Psi_2 = \sqrt{3}\;J_0(\kr r_2)=0.52$ a.u., \ $\Psi_3 =
\sqrt{4}\;J_0(\kr r_3)=-0.5$ a.u.

\begin{figure}[!h]
\begin{center}
 \scalebox{2.2}{\includegraphics{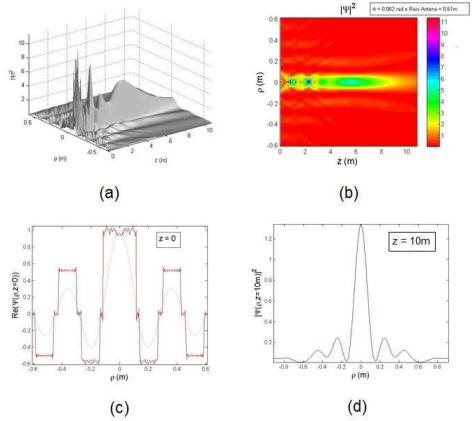}}
\end{center}
\caption{{(Color online) Field emanated by Prototype 2. \ See the text.}} \label{figure7}
\end{figure}

\h Figures \ref{figure7} describe the field emanated by Prototype 2.

\h Here the idea is that of increasing the field intensity in the slits (except for the central
aperture), since in this way the spot radius does not vary, but the beam intensity distribution
at $\rho=0$ gets more homogeneous than for Prototype 1. \ One can see, moreover, that the spot
intensity at $z=10$ m gets substantially improved.

\

{\bf Prototype no. 3:} --- Once more, let us keep unaltered the antenna dimensions, and modify only the value of the
uniform fields illuminating the slits. \ Namely, apply the same amplitude magnitude (that is, {\em the same
intensity\/}) to all the slits!, changing only its sign, which will alternately be positive or
negative when passing from one slit to the other: In other words, let us change only the phase [by the trivial
quantity $\pi$] when going from one slit to the next one. \ Numerically, we shall have: $\Psi_0 = 1$ a.u., \ $\Psi_1 = -1$ a.u., \ $\Psi_2 = 1$ a.u., \ $\Psi_3 = -1$ a.u.

\begin{figure}[!h]
\begin{center}
 \scalebox{2.75}{\includegraphics{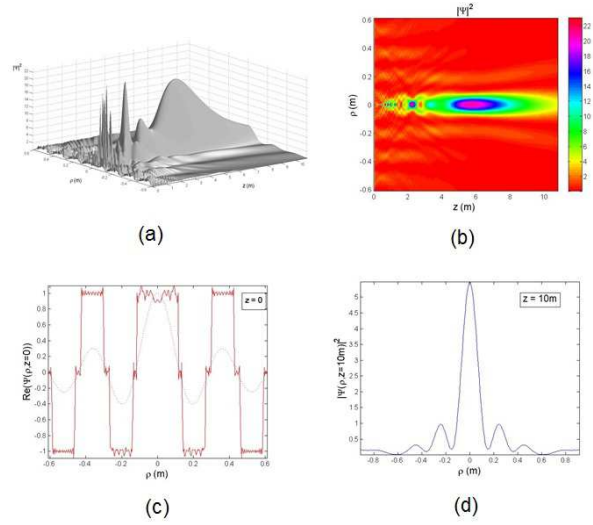}}
\end{center}
\caption{{(Color online) Field emanated by Prototype 3. \ See the text.}} \label{figure8}
\end{figure}

\h Figures \ref{figure8} describe the field emanated by Prototype 3.

\begin{figure}[!h]
\begin{center}
 \scalebox{3.0}{\includegraphics{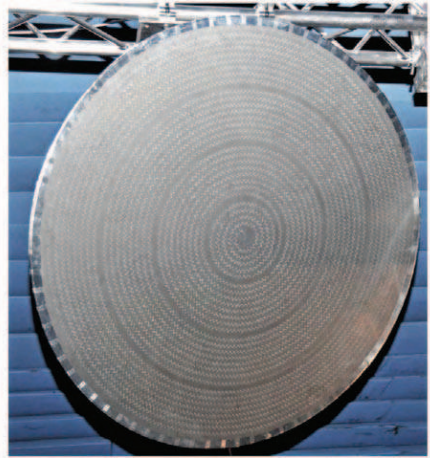}}
\end{center}
\caption{{(Color online) The antenna concretely constructed starting from proposals as the above
ones\cite{MRBunpub1,Antenna3Archives,mrb2,PatentII}; its diameter being 120 cm.  More details in Refs.\cite{Polaris,IEEEbalma}. {\em Photo courtesy of M.Balma.}}}
\label{figureAntenna}
\end{figure}

Notwithstanding the homogeneous intensity distribution in the slits, the emanated field rapidly becomes a beam with a field
concentrated around $\rho=0$. \ This is just due to the alternate phase change (every time of the quantity $\pi$) when
passing from a slit to the next one.

\

\

\h It is worth mentioning that an antenna based on the above proposals\cite{PatentII,Antenna3Archives,mrb2,MRBunpub1} has been {\em concretely constructed}
(see Figure \ref{figureAntenna})  and efficiently applied for obtaining for instance a spot of about 10 cm at a distance
of about 10 m, as a significant help for detecting, e.g., buried explosive mines at a safe distance\cite{Polaris,IEEEbalma,PatentII}.

\

\h Let us discuss, and clarify, the structure of the real antenna in Figure \ref{figureAntenna}). It was a product of
our approach\cite{MRBunpub1,Antenna3Archives,mrb2,PatentII}, presented above and exemplified by Figure \ref{figure5}. Within each interval, however, the annular slit was optimized\cite{Polaris,IEEEbalma} and transformed into a long, spiraling slit. Such spirals were recognized by us to be simply of the Archimede type\cite{MRreport3,PatentII}.

\h In fact, the antenna in Figure \ref{figureAntenna} (formed by four sets of spirals) does indicate by itself that each one
of those sets corresponds to a single Archimede's spiral: Indeed, by drawing any straight lines passing through the
origin, it can be seen that each straight line intersects the spiral at points separated, one from the other, by a constant distance. And this is the fundamental characteristic of an Archimede spiral.

More precisely, the rings constituting the discretized and optimized antenna appear to be formed by slits (we can call them slots), placed along Archimede spirals which go from a root of the Bessel function  $J_0(\kr \rho)$ to the next one. If we recall that that an Archimede spiral in polar coordinates gets the simple form  \ $\rho = a + b\phi$, \ with $a$ and $b$ constant, then
the previous observation means that we shall have:

\bb \rho = \rho_{0i} + b_i \phi \ \ \ \rm{for} \ \ \ \rho_{0i} \leq \rho \leq \rho_i - \delta/2
\label{r1}
\ee

where quantities $\rho_i$ are the mentioned roots of the Bessel function [given, therefore, by the equation
$J_0(\kr \rho_i) = 0$], \ it being \ $\rho_{01} \equiv \rho_0$; \  while the subsequent values of $\rho_{0i}$ will obey the relation \  $\rho_{0i} = \rho_{i-1} + \delta/2$. \ When changing antenna, it will vary only the value of $\de$.

\h In general, the values of quantities $b_i$ in Eq.(\ref{r1}) are given by

\bb b_i = \frac{(\rho_i - \de/2) - (\rho_{i-1} + \de/2)}{2m_i\pi} = \frac{\rho_i -\rho_{i-1} -
\de}{2m_i\pi} \ee

where we called $m_i$ the number of turns of the spiral in the interval $\rho_{0i} \leq \rho \leq \rho_i - \delta/2$.

The spirals, covering each of the intervals \ $\rho_i - \de/2 \leq \rho \leq \rho_i + \de/2$, correspond therefore
to the quantities

\bb \rho = \rho'_{0i} + c_i\phi \ee

where

\bb \rho'_{0i} = \rho_i - \de/2  \ee

and

\bb c_i = \frac{(\rho_i + \de/2) - (\rho_i - \de/2)}{2\pi} = \frac{\de}{2\pi} \ . \ee

If the field is represented by a Bessel-beam field ---in which case it can be simply regarded as a linearly polarized
field---, then amplitude and phase of each slot can be immediately furnished by the Bessel function.

\h The antenna in Figure ref{figureAntenna} constitutes the first concrete example
of application of the NDWs in the sector of electromagnetic waves, the unique previous
experiment having been the known one by Ranfagni, Mugnai and Ruggeri employing microwaves
(but with a demonstrative purpose, and by a totally different technique. A couple
of experiments had been also performed in Optics; besides the Lu et al. ones, in
Acoustics). Incidentally, our theoretical predictions, and concrete suggestions, were
confirmed by the numerical simulations performed in \cite{Polaris} and \cite{IEEEbalma}.
One should not forget, moreover, that our antenna using NDWs produces a 10 cm spot till
at least 10 m of distance.

\begin{figure}[!h]
\begin{center}
 \scalebox{3.25}{\includegraphics{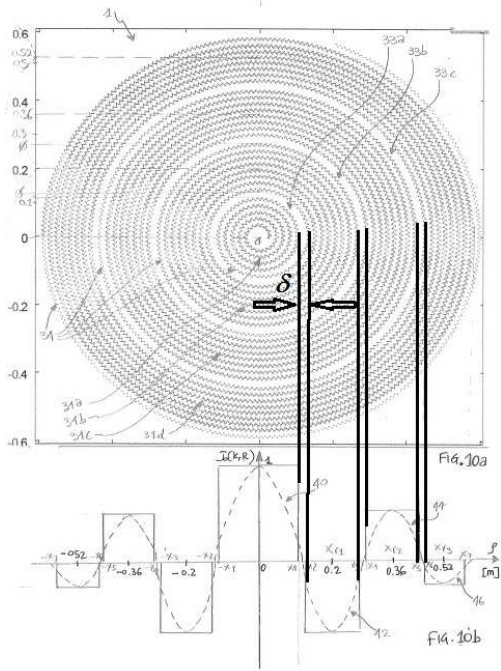}}
\end{center}
\caption{{The antenna in the previous figure has been concretely constructed also by exploiting our
approach\cite{MRBunpub1,Antenna3Archives,mrb2,PatentII}, presented in this Section and exemplified by Figure \ref{figure5}. Within each interval, the position of the annular slits was
optimized\cite{Polaris,IEEEbalma}, yielding actually spiraling slits: Such spirals were recognized by us to be simply
of the Archimede type\cite{PatentII}, as explained above.}}
\label{figureSpirali}
\end{figure}

\h Further remarks: (i) the point at which the spirals start, corresponding to the
value of $\rho$ that we called $\rho_0 = \rho_{01}$, is determined a priori, since is
given by the shape of the adopted antenna (and it is not at all the origin O); \ (ii)
the quantity $\delta$ appearing above is nothing but the distance between the rectangles
adopted to schematize and discretize (``by steps") the Bessel function (see Figure \ref{figureSpirali}).
We had chosen it as a constant, and it remained so even in the optimized
antenna. However, it is chosen a priori; and it might change from a couple of rectangles to the next one,
becoming a variable $\delta_irm$; \ (iii) one should not forget that each set of slots is a single
Archimede spiral (as we specified above).

\

\

{\bf Some more comments:} --- We have mentioned above, in particular, the purpose of remote sensing: For instance,
of the remote detection of the presence of a buried object. Several approaches may be suggested:

\

1) In the case of microwaves, one has to produce suitable NDWs.
To such an aim, one can suggest their generation:

1a) either by annular slits,

1b) or by a parabolic mirror, with a source slightly displaced
from the focal position along the principal axis,

1c) or by circular electric currents; or, rather, by electrically
feeding discrete conductive elements, located along rings.

\

Let us add, however [we are forgetting in this work about the ``Frozen Waves"], that one could deal also with:

\

2) the possibility of producing (by {\em acoustic} transducers) a {\em sonic
bullet,} both for remote detection, and even to the aim of producing by mere compression the
explosion of the buried object, in case it is a mine. Let us recall that
---in terms of suitable superpositions of equal-frequency Bessel beams--- we developed theoretical
methods to obtain analytic expression for NDWs even in absorbing media.  Anyway, we shall deal in other papers with
the (acoustic) case of generators of Non-Diffracting ultrasonic waves.

\

\

\section{Antenna composed by a Parabolic Reflector and a Spherical Wave Source slightly
Displaced w.r.t. the Focus of the Paraboloid}

We want now consider just the alternative possibility, based on the properties of the parabolic mirrors,
whose equation is \ $z = a\rho^2$, to use paraboloids as a source of Non-Diffracting microwaves.

\ It is well known that any ray passing through the focus (at
$z_f = 1/4a$) of a paraboloid is reflected parallel to the reflector axis $z$. If we move our spherical wave source
away from the focus (on the $z$-axis and in the positive $z$ direction),
the rays striking the reflector will cross the $z$-axis in a position that depends on the incidence
point: Namely, the reflected rays will converge on a focal segment (``extended focus").

\begin{figure}[!h]
\begin{center}
 \scalebox{3.0}{\includegraphics{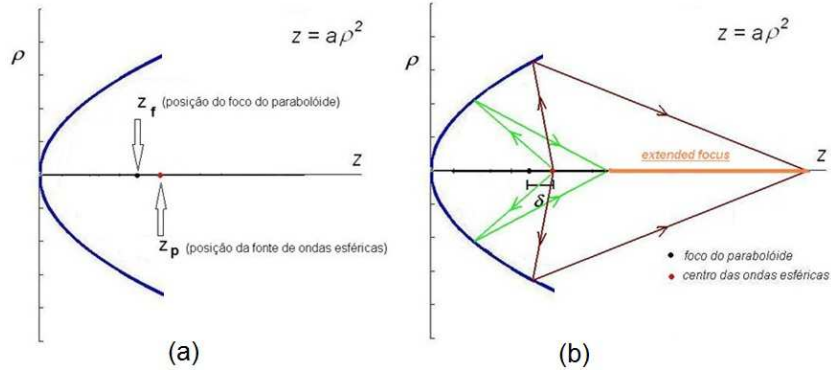}}
\end{center}
\caption{(Color online) Parabolic Reflector with a spherical wave source --assumed to emit, e.g., 30 GHz waves-- located along the $z$-axis but at a position shifted
with respect to the focus of the Paraboloid: See the text.} \label{figure9}
\end{figure}

\h Figure \ref{figure9} is self-explicative.  The point $(\rho=0,\ z=z_f)$ is the paraboloid focus.
The point $(\rho=0, \ z=z_p)$ is situated at the right of the focus. We shall choose a frequency og 30 GHz.

Geometric optics tells us that a ray, starting from $z_p$ and incident on the paraboloid at the point $(\rho',z')$, after
having been reflected will meet the $z$-axis at a distance $D$ from the reflector vertex given by

\bb D = \frac{4az'^2 + 2z' + z _p}{4az_p - 1} \; ;   \ee

this equation yields the minimum and maximum values of $D$ as functions of $z'$. Assume that $0\leq z' \leq z_p$. It is
easy to show\cite{MRBunpub2,MRunpub} that

\bb D_{\minrm} \ug \frac{z_p}{4az_p - 1} \; , \ee

while the $D_{\maxrm}$ corresponding to $z'=z_p$ is

\bb D_{\maxrm} \ug \frac{4az_p^2 + 3z_p}{4az_p - 1} \ug z_p + 4D_{\minrm} \; . \ee

One therefore gets that the ``focal width" $Z_{\rm focalwidth}$ will be

\bb Z_{\rm focalwidth} \ug D_{\maxrm} - D_{\minrm} \ug z_p + 3D_{\minrm} = z_p + \frac{3 z_p}{4az_p - 1} \; .  \ee

In many situations one has $z_p<<D_{\minrm}$, so that:\cite{MRBunpub2,MRunpub}

\bb Z_{\rm focalwidth} \approx \frac{3 z_p}{4az_p - 1} \ug 3D_{\minrm} \; . \ee

\h A basic characteristic of ND beams is possessing not a point-like focus, but just an extended focus
(or focal segment): Therefore, one can intuitively expect the present setup to furnish a beam of non-diffracting (ND) type;
even if all these considerations are based on geometric optics, which implies for instance (as a necessary, but not sufficient, condition) that the paraboloid size is much bigger than $\lambda$. If we suppose the emanated field to be non-diffracting,
this will take place in the interval $D_{\minrm}<z<D_{\maxrm}$. \ When it is, moreover, $z_p<<D_{\minrm}$, that interval
becomes \ $D_{\minrm} < z < 4 D_{\minrm}$, \ with \ $D_{\minrm} = z_p/(4az_p - 1)$: \ And our ``ND beam" will start to exist at
the distance $D_{\minrm}$ from the antenna, and will go on existing till the {\em triple} of such a distance.

\h Before going on, let us repeat that the present paper reports about work performed by us during 2011, and 2012. \
Subsequently, however, further analogous work on the use of paraboloids as antennas has been done by other friends
of ours: So that the interested reader might find more results in Refs.\cite{TesiMariana,MMLunpub}.

\begin{figure}[!h]
\begin{center}
 \scalebox{3.1}{\includegraphics{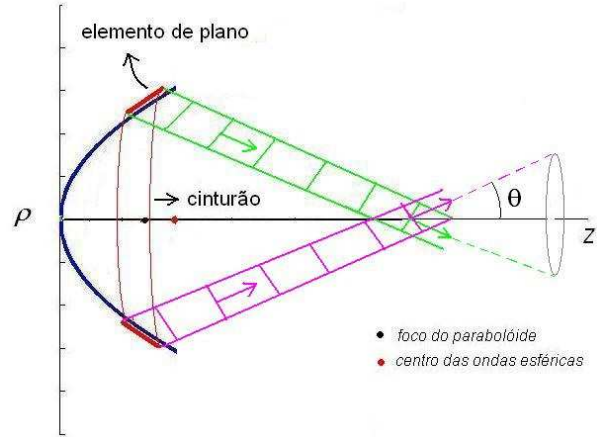}}
\end{center}
\caption{(Color online) Discretization of the paraboloid into circular rings (see {\em cinturão,} in the Figure) , with a successive discretization of the rings themselves into flat elements. The figure also shows the portions of wave reflected by two of such flat elements. See the text.} \label{figure10}
\end{figure}

\h When a ND beam is generated in the interval $D_{\minrm} < z < 3 D_{\minrm}$, the interesting problem arises of
knowing the evolution of its transverse intensity during propagation\cite{MRBunpub2,MRunpub,TesiMariana,MMLunpub}.
Some information can
be obtained on the basis of the following considerations, which imply however that the parabolic reflector be much,
much larger than $\lambda$ (a situation certainly true in Optics, but not necessarily for microwaves). \ Let us imagine
of dividing the reflector into circular rings (see {\em cinturão,} in Figure \ref{figure10}) of sufficiently small width; and afterward
of splitting each ring into flat elements, sufficiently small as well but with sizes much larger than $\lambda$.
\ One can then think that each portion of the spherical
wave (originated at $\rho=0,z=z_p$), when incident on one of such flat elements, is reflected in the form of a portion of planewave (having the size of the considered flat element) and travels without appreciable diffraction till the $z$-axis.  Thus, at each point of the $z$-axis it will arrive a set of small portions of plane waves reflected by the corresponding circular ring.\footnote{One should remember that each point $\rho',z'=a\rho'^2$ of the reflector corresponds to a precise point of the $z$-axis}. For
symmetry reasons, the wave-vectors of the said set of small portions of planewave will stay on the surface of a cone with
angle $2\theta$. See Figure \ref{figure10}. Such a superposition will give rise to a Bessel beam with axicon angle
$\theta$, that will propagate along a short interval of $z$, being then replaced by the Bessel beam coming from the next
circular ring.  This will repeat itself till the distance $D_{\maxrm}$.

\h Still from purely geometric considerations, is follows also that:\cite{MRBunpub2,MRunpub}

\bb \sin\theta \ug \frac{1}{\sqrt{\left(\frac{z-a\eta^2}{\eta}\right)^2 + 1}} \ee

where

\bb \eta \ug \sqrt{-\frac{1}{4a^2} + \frac{1}{4a^2}\sqrt{1 - 4a(z_p-(4az_p-1)z)}} \ . \ee

\h This allows us to state, even if in an approximate way, that all along the ``extended focus", and for
point near the $z$-axis, the field will be proportional to a Bessel function with a {\em (variable)} axicon
angle $\theta$. That is:

\bb \Psi(\rho,z,t) \propto e^{-i\om t}J_0\left(\frac{\om}{c}\sin\theta\,\rho\right) \label{bb2} \; \ee

where $\theta$ is a complicate function of $z$, as we have just seen above; and we still choose a 30 GHz frequency. \ In the last equation, the proportionality
symbol means ``except for a multiplicative function depending on $z$ and $\rho$:" \ It is this function that determines
the {\em magnitude} of the wave.  Therefore, the approximate expression (\ref{bb2}) does not yield the varying
intensity of the field during propagation, but only its transverse behavior.

\

\h {\bf --- Let us now propose the following antenna:}

\

\begin{itemize}
    \item Position of the focus: $z = z_f = 0.5\,$m \ (so that $a=1/4z_f=0.5\,m^{-1}$);
    \item Position of the spherical wave (30 GHz) source: $z = z_p = 0.525\,$m \ (shifted 2.5 cm from the focus
     in the positive $z$ direction).
\end{itemize}

\

\begin{figure}[!h]
\begin{center}
 \scalebox{2.7}{\includegraphics{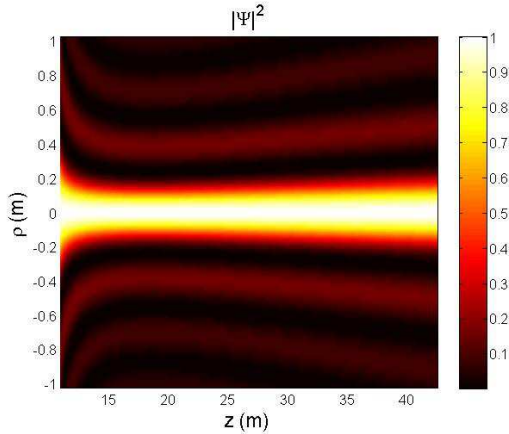}}
\end{center}
\caption{(Color online) Orthogonal, transverse projection of the field emanated by the parabolic reflector, when assuming: Position of the focus, $z = z_f = 0.5\,$m \ (so that $a=1/4z_f=0.5\,m^{-1}$); \ Position of the spherical wave (30 GHz) source: $z = z_p = 0.525\,$m \ (shifted 2.5 cm from the focus in the positive $z$ direction). See the text. This Figure shows intensity and evolution of the emanated field, according to Eq.(\ref{bb2}). [Let us recall that this Figure gives us information on the evolution,
starting from the the aperture, of the field transverse behavior, but not the exact intensity of the produced beam].
} \label{figure11}
\end{figure}

Assuming for the paraboloid $0 \leq z \leq z_p$, its ``mouth" will have a radius $R = 1.02\,$m. \ Such a configuration
furnishes: $D_{\minrm}=11.02\,$m \ \ \ $D_{\maxrm}=42.52\,$m; \ and one does expect that along the segment
$D_{\minrm} < z < D_{\maxrm}$ a ND beam is formed. Figure \ref{figure11} shows intensity and evolution of its
transverse shape, according to Eq.(\ref{bb2}). [Let us repeat that such a Figure gives us information on the evolution,
starting from the the aperture, of the transverse behavior, but not yet the exact intensity of the produced beam].

One can see that this beam possesses an initial spot with radius $\Delta\rho(z=0)=50\,$cm, if we choose as spot
radius the value of $\rho$ where it occurs the first zero of the field intensity. Subsequently, the spot radius
changes during propagation, diminishing till the approximate value of 24 cm at the point $z \approx 18\,$m, beyond which
it starts to increase again, ending with a value of about 31 cm at $z=45\,$m. Roughly speaking, we can call such a
beam a ``ND beam", since its spot radius does not exceed the initial value even after having traveled for long
distances.

\begin{figure}[!h]
\begin{center}
 \scalebox{1.3}{\includegraphics{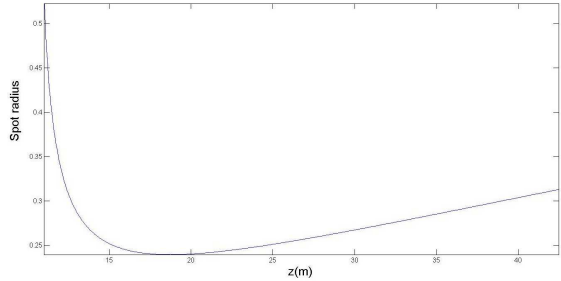}}
\end{center}
\caption{(Color online) Variation of the spot radius for the produced beam, as a function of the distance from the
aperture. See the text.} \label{figure12}
\end{figure}

\h Figure \ref{figure12} shows the variation of the spot radius as a function of the distance from the aperture.
It represents a ND beam generated along more than 30 m; but such a value should not be taken literally since we
adopted a model developed ``in Optics", that is, when the reflectors can easily have sizes much bigger than $\lambda$.

\

\

\h {\bf --- Some further examples and results:}

\

\begin{figure}[!h]
\begin{center}
 \scalebox{2.7}{\includegraphics{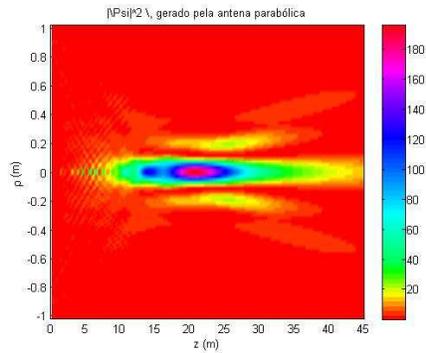}}
\end{center}
\caption{(Color online) Orthogonal, transverse projection of the field emanated by an identical parabolic reflector, when the source of spherical waves (with
frequency 30 GHz) is slightly displaced w.r.t. the focus, so that the initial spot of the generated field is of about
10 cm.  One can see that such a spot, even with a non constant intensity, goes on keeping its size ---apart from a {\em slight} enlargement---  till a distance of almost 50 m \ (the underlying analysis has still been scalar). \ The following two Figures show, for comparison, the behavior of analogous
gaussian beams.
} \label{figure13}
\end{figure}

Let us mention some further results. Assume again, for the paraboloid, the focal position
at 0.5 m, and a radius at its mouth of 1 m.  We can locate a source of the spherical waves, with frequency
30 GHz, by shifting it a little bit w.r.t. the reflector focus, in order that the initial spot of the generated field
is of about 10 cm. \ Figure \ref{figure13} shows that such a spot, even with a non constant intensity, goes on keeping its size
---apart from a {\em slight} enlargement---  till a distance of almost 50 m. The underlying analysis has still been scalar. \ The following two Figures show, for comparison, the behavior of analogous gaussian beams: They, when starting with the same
spot size, get a field-depth of less than 6 m.

\begin{figure}[!h]
\begin{center}
 \scalebox{2.7}{\includegraphics{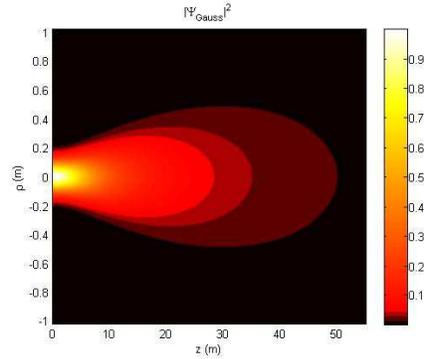}}
\end{center}
\caption{(Color online) For comparison with the previous Figure, notice as a gaussian beam with the same initial
spot possesses a depth of field of less than 6 m. \ The present Figure refers to a gaussian beam with focal point
at $z=0$.
} \label{figure14}
\end{figure}

\begin{figure}[!h]
\begin{center}
 \scalebox{2.7}{\includegraphics{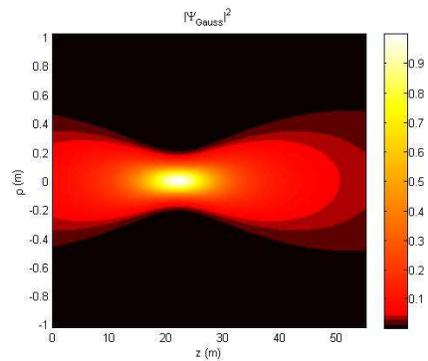}}
\end{center}
\caption{(Color online) The same as in the previous Figure, when the gaussian beam focal point is located
at $z=22$ m. \ Again, the field depth is of less than 6 m, instead of almost 50 m.
} \label{figure15}
\end{figure}

\

\

\h {\bf --- A few final considerations:}

Various other antennas are possible, varying the values of $z_f$ and $z_p$; besides modifying the antenna size
itself.

\begin{figure}[!h]
\begin{center}
 \scalebox{1.5}{\includegraphics{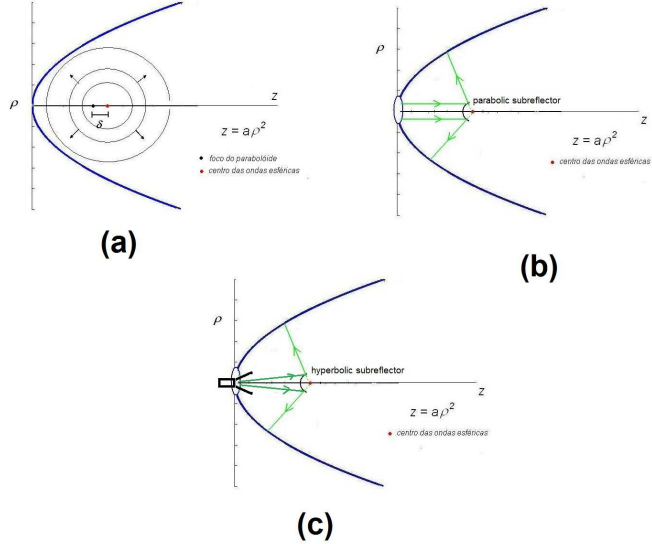}}
\end{center}
\caption{(Color online) The source of spherical waves, located at $z = z_p$, can be approximated by a spherical, or hyperbolic, mirror ({\em sub-reflector}) reflecting the radiation coming from a suitable hole surrounding the vertex. \ See the text for more details.}
\label{figure16}
\end{figure}

\h But let us spend, rather, a few words about the spherical wave source (cf. Figure \ref{figure16}a)
to be located at $z=z_p$. Of course,
it cannot be large, to avoid blocking a considerable part of the wave reflected by the paraboloid. We may just
set forth one suggestion: One can think to open a circular hole around the paraboloid vertex, and send through it a
collimated beam parallel to the $z$-axis. At position $z_p$ we can then put a spherical reflector (see Figure \ref{figure16}b),
so that the parallel rays (coming from the hole created around the paraboloid vertex) are reflected towards the interior of the paraboloid
itself. These rays would behave as if emitted from the point $z_p$.

\h Of course, the hole radius must be smaller than or equal to the
reflecting spherical mirror's, both radii having to be as small as possible; at last, the beam entering the hole {\em must}
remain parallel to $z$ as much as possible. \ A problem which remains to be taken into account is that the intensity
of the rays reflected by the spherical mirror is not isotropic, lowering its value for increasing reflection angle.

\

\

\section{Acknowledgements}

The present work has been partially supported by FAPESP, CAPES and CNPq (Brazil), and by INFN (Italy).
For kind collaboration or useful discussions during the past several years, the authors are grateful to
I.A.Besieris, R.Bonifacio, C.Castro, R.Chiao, C.Conti, A.Friberg, D.Faccio, F.Fontana, E.Giannetto,
P.Hawkes, R.Grunwald, G.Maccarini, M.Mattiuzzi, C.Meroni, P.Milonni, M.Novello, S.Paleari, C.Papa, P.Riva,
J.L.Prego-Borges, P.Saari, A.Santambrogio, A.Shaarawi, M.Tygel, A.Utkin, R.Ziolkowski, and particularly to M.Balma,
D.Campbell, H.E.Hern\'andez-Figueroa, and M.Mojahedi.

\

\

\

\end{document}